\documentclass[floatfix,twocolumn,prl,showpacs,amsmath]{revtex4}
\usepackage{graphicx}
\usepackage{subfigure}
\usepackage{amssymb,amsmath}
\usepackage{mathrsfs,dsfont}

\begin{document}
\title{Self-Diffusiophoresis in the Advection Dominated Regime}
\author{Gareth P. Alexander}
\affiliation{Department of Physics \& Astronomy, University of Pennsylvania, 209 South 33rd Street, Philadelphia PA 19104, U.S.A.}
\author{Andrea J. Liu}
\affiliation{Department of Physics \& Astronomy, University of Pennsylvania, 209 South 33rd Street, Philadelphia PA 19104, U.S.A.}
\date{\today}

\begin{abstract}
In both biological and artificial systems, concentration gradients can serve as a convenient mechanism for manipulating particles and generating motility. Particles that interact with a solute will move along its gradient; if they themselves generate the gradient, this mechanism provides a means of self-propulsion. We consider a version of this type of motility appropriate to certain biological systems where polymeric filaments provide the concentration gradient. As the filament diffusion is small, this corresponds to a regime of large P\'eclet number where the motion is dominated by the effects of fluid advection. The nature of such concentration-gradient-driven motion in the advective regime differs in certain fundamental respects from the same process at low P\'eclet number. In particular, we show that out of four broad scenarios of steady state motion at low P\'eclet number, only two remain viable in the strongly advecting limit.  
\end{abstract}
\maketitle

Certain motile biological objects, such as the bacterium {\it Listeria monocytogenes} and replicating chromosomes in asymmetric bacteria such as {\it Caulobacter crescentus} and {\it Vibrio cholerae}, generate their propulsion by polymerizing or depolymerizing protein filaments~\cite{cameron99,plastino05,fogel06}. This establishes a concentration gradient in the protein so that there are more filaments on one side of the object than the other. In the case of {\it Listeria}, the activation of the Arp2/3 protein complex at the cell posterior promotes the polymerization of actin and results in the formation of a distinctive comet tail, a region of high actin concentration immediately behind the moving bacterium~\cite{cameron99}. Conversely, in {\it Caulobacter} and {\it Vibrio} the chromosome stimulates the disassembly of filaments of ParA so that the concentration of ParA is higher in front than it is behind the chromosome as it translocates~\cite{shebelut}.  

A colloid that maintains an asymmetric concentration of solute around it will propel itself through a fluid with a well-defined velocity.  This phenomenon, known as self-diffusiophoresis, has been exploited as a means of propulsion of micro- or nano-swimmers~\cite{paxton04,golestanian05,golestanian07,howse07,golestanian09,popescu09,popescu10,sabass10}.  In a typical example, the concentration gradient is controlled by an active region on the colloid that catalyzes a chemical reaction, leading to more product and less reactant near the active region than on the far side of the colloid. An interaction between the colloid and solute sets up fluid flow within a thin boundary layer with a slip velocity at the edge of this layer proportional to the concentration gradient~\cite{anderson89,golestanian07}. This local concentration gradient is in turn determined by a combination of diffusive and advective fluxes, whose relative importance is measured by the P\'eclet number $\text{Pe} = Ua/D$, where $U$ and $a$ are a characteristic velocity and length scale of the colloid, respectively, and $D$ is the solute diffusion constant.  In experimental realizations considered to date (such as the decomposition of H$_2$O$_2$~\cite{howse07}), the solute has a large diffusion constant and Pe is small ($\sim 10^{-3}$) so that the concentration gradient is effectively set by diffusion alone.  Similarly, theoretical treatments have considered only the limit Pe=$0$~\cite{golestanian07,golestanian09,popescu09,popescu10,sabass10}.  In this limit it is known that only two conditions are needed to achieve motility:  the motile object must be able to maintain an asymmetric solute distribution in steady state, and there must be a net interaction between the solute and the object.  However, in the biological examples of interest, the relevant solutes are filamentous protein assemblies with effectively vanishing diffusion constants and it is not known whether the same conditions apply for motility.  

In this Letter, we determine the general properties of propulsion in a self-generated concentration gradient in the limit of infinite P\'eclet number.  We find that advective transport through the boundary layer leads to significant differences in the conditions needed to sustain steady state motion and in the scaling of the speed with surface activity, as compared to the Pe=$0$ limit.  Some of our results are consistent with known observations of polymerization- or depolymerization-driven motility involving filamentous proteins, and the remainder can potentially be tested in such systems.

\begin{figure}
\centering
\includegraphics[width=.49\textwidth]{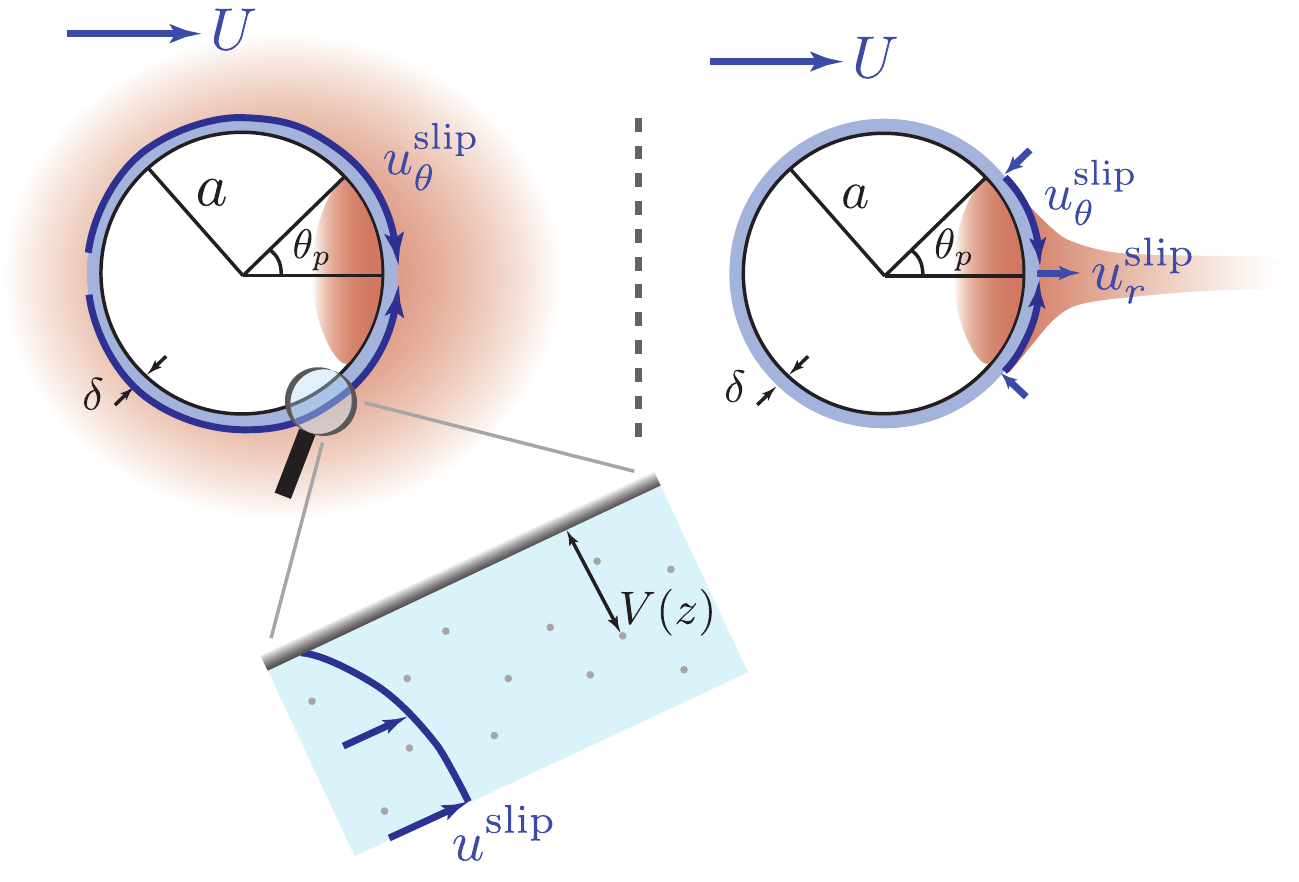}
\caption{(Color online) Schematic illustration of self-diffusiophoresis for a colloid at low (left) and high (right) P\'eclet number indicating the form of the slip velocity. The magnified region shows the nature of the approximation used in the boundary layer analysis.}
\label{fig:schematic}
\end{figure}

We consider the idealized minimal model of Ref.~\cite{golestanian07}, in which a spherical particle of radius $a$ produces solute from an active patch on its surface at a rate $\alpha$, shown schematically in Fig.~\ref{fig:schematic}. We restrict our attention to axisymmetric, steady state motion, viewed from the rest frame of the particle. The particle interacts with the solute through a potential $V(r)$ with a range $\delta \ll a$ that is small compared to the particle size;  this separation of length scales allows the effect of the interaction to be determined through a boundary layer analysis that neglects surface curvature on the scale $\delta$. Consequently, the flow within the boundary layer is well approximated by the solution of the Stokes equations 
\begin{equation}
{\bf 0} = - \nabla p + \mu \nabla^2 {\bf u} - c \nabla V , \qquad \nabla \cdot {\bf u} = 0 ,
\label{eq:Stokes}
\end{equation}
where $\mu$ is the fluid viscosity and $c$ the solute concentration, in the upper half space ($z\geq 0$) that approximates the local environment near the particle surface, as seen on the scale $\delta$. The resultant flow is tangential to the surface with a limiting value, known as the slip velocity, at the outer edge of the boundary layer of  
\begin{equation}
{\bf u}^{\text{slip}}({\bf x}) = - \frac{1}{\mu} \int_0^{\delta} \text{d}z \, \frac{1}{2} z^2 \, \nabla_{\parallel} c(z,{\bf x}) \, \partial_z V(z) , 
\label{eq:slip_general}
\end{equation}
where $\nabla_{\parallel}$ denotes gradients projected into the tangent plane of the surface. When Pe is small and the solute diffuses rapidly it is a good approximation to assume that the concentration maintains a local equilibrium with a Boltzmann distribution $c(z,x) = c(\delta,x) \, {\rm e}^{-V(z)/k_BT}$, which reduces the slip velocity to ${\bf u}^{\text{slip}} = m^D \nabla_{\parallel} c$ where $m^D = \tfrac{k_BT}{\mu} \int_0^{\delta} \text{d}z \, z [1 - {\rm e}^{-V(z)/k_BT}]$ is a diffusiophoretic mobility~\cite{anderson89}. 

At arbitrary Pe, the solute concentration field obeys a conservation equation outside the boundary layer 
\begin{equation}
\partial_t c + {\bf u} \cdot \nabla c - D \nabla^2 c = 0 ,
\label{eq:conserve_solute}
\end{equation}
together with the boundary condition 
\begin{equation}
\int_{r=a+\delta} \bigl[ u_r^{\text{slip}}c - D \partial_r c \bigr] = \int_{r=a} \alpha ,
\label{eq:solute_boundary_condition}
\end{equation}
expressing that in steady state the rate of production of solute at the particle surface must equal the rate at which it is transported across the boundary layer and into the bulk fluid. At small Pe, the advective terms in Eqs. \eqref{eq:conserve_solute} and \eqref{eq:solute_boundary_condition} may be neglected, while for thin boundary layers the integral boundary condition is replaced with a pointwise equality $-D \partial_r c = \alpha$ \cite{golestanian07}, which yields an unique solution for the concentration field. In the opposite limit of large Pe the solute is still conserved and Eqs. \eqref{eq:conserve_solute} and \eqref{eq:solute_boundary_condition} still apply.  However, if diffusion is negligible the solute can only be transported away from the particle surface if it is advected by a radial flow. This radial slip velocity, neglected in the tangential projection of Eq. \eqref{eq:slip_general}, is therefore crucial to retain in the advective regime. 

However, the presence of a radial slip does not invalidate Eq. \eqref{eq:slip_general}, as this is only an asymptotic formula that neglects surface curvature on the scale $\delta$. Dimensional analysis of the fluid continuity equation shows that the radial slip is typically smaller than the tangential slip by a factor $\delta/a$, so that its neglect is based on the short range of the interaction of the solute with the particle surface.   Since this range remains short at high Pe, Eq.~\eqref{eq:slip_general} will remain a good approximation, with one caveat, namely, that any tangential flow around a spherical body must vanish somewhere (and the sum of the indices of all the zeros must be $2$ -- Poincar\'e-Hopf theorem) and in the vicinity of any such point the neglect of the radial slip is not justified. 

Away from these zeros the slip will continue to be approximated by the tangential boundary layer flow, which should behave as in Eq.~\ref{eq:slip_general}:
\begin{equation} 
u_{\theta}^{\text{slip}}(\theta) = \frac{m^A}{a+\delta} \, \partial_{\theta} c(a+\delta,\theta) ,
\label{eq:utheta_slip}
\end{equation}
in spherical coordinates, where $m^A$ is an `advective mobility.'  A simple but significant observation coming from this relation is that there is no tangential flow where there are no gradients. At low Pe diffusion spreads the solute throughout space, at least close to the particle. However, if the solute does not diffuse then it will only be found where it is produced, or has been transported by the fluid. Thus if we consider an active patch, covering the portion $0 \leq \theta \leq \theta_p$ of the sphere, the vanishing of the solute concentration everywhere outside the patch implies that the tangential slip velocity also vanishes there. Fluid continuity then requires that the tangential flow within the patch region is supplied by a radial influx at the edge $\theta = \theta_p$ and subsequently escapes via a radial outflux localized near some internal point, $\theta = 0$ by axisymmetry. These considerations determine the qualitative features of the flow at high P\'eclet number. 

To be more quantitative we take this picture as an idealization, so that there is radial flow only at $\theta = \theta_p$ and $\theta = 0$, and purely tangential slip elsewhere. Fluid continuity constrains this tangential flow so that $\sin(\theta)u_{\theta}^{\text{slip}}$ is a constant, which we write as $-\lambda$ with $\lambda>0$. Eq. \eqref{eq:utheta_slip} then allows us to solve for the solute concentration profile at the edge of the boundary layer 
\begin{equation}
c(a+\delta,\theta) = \frac{\lambda (a+\delta)}{m^A} \, \text{ln} \, \biggl[ \frac{\tan(\tfrac{1}{2}\theta_p)}{\tan(\tfrac{1}{2}\theta)} \biggr] . 
\label{eq:solute_profile}
\end{equation}
The singular behavior as $\theta \rightarrow 0$ is an artifact of neglecting the radial slip.   We regularize this by cutting off the solution at some small angle $\theta_q$ within which the concentration assumes a constant value. 

The net radial outflux from the boundary layer is equal to the net tangential influx across a cylindrical surface spanning the thickness of the boundary layer at any $\theta_q < \theta < \theta_p$, which for thin boundary layers we approximate as $-2\pi a \delta \beta \sin(\theta)u_{\theta}^{\text{slip}}$, where $\beta$ can be expected to be an ${\cal O}(1)$ constant coming from the average over the boundary layer. Thus the radial fluid outflux is given by 
\begin{equation}
\int_0^{\theta_q} \text{d}\theta \, \sin(\theta) u_r^{\text{slip}} = \frac{\beta a \delta}{(a+\delta)^2} \lambda ,
\label{eq:radial_outflux_2}
\end{equation}
and is small compared to the tangential flow by a factor $\delta/a$ as expected. 

\begin{figure}
\centering
\includegraphics[width=.48\textwidth]{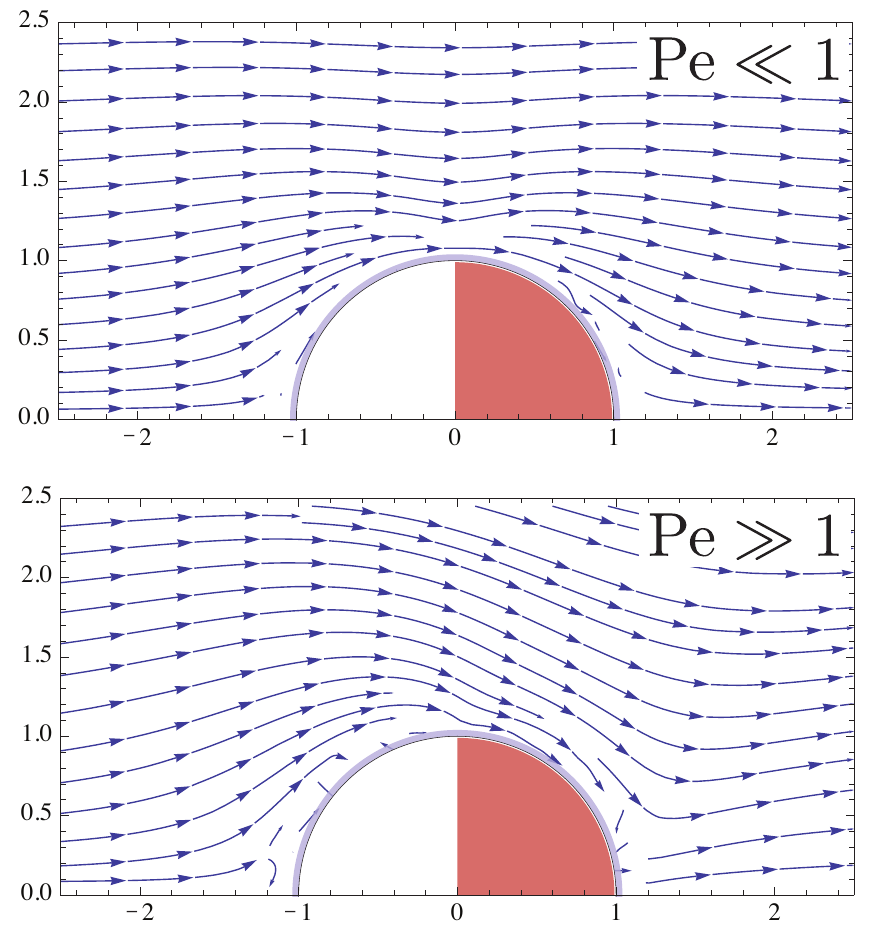}
\caption{(Color online) The flow field of a self-diffusiophoretic Janus particle at small (top) and large (bottom) Pe. The flow is shown scaled in units of the translational speed $U$ and distances are scaled in units of the particle radius $a$. Note that the flow exhibits a fore-aft symmetry at low Pe, which is broken when the P\'eclet number becomes large.}
\label{fig:flow}
\end{figure}

At low Pe, the radial flow from the boundary layer can be neglected because $\delta/a \ll 1$.  However, it is a feature of the high Pe regime that this radial outflux provides the sole mechanism for the transport of solute produced at the surface through the boundary layer and into the bulk fluid. This transport is given by Eq. \eqref{eq:solute_boundary_condition} with diffusion neglected and leads, for uniform activity $\alpha(\theta) = \bar{\alpha}$ within the active patch $0 \leq \theta \leq \theta_p$, to an expression for the tangential flux $\lambda$ 
\begin{equation}
\lambda^2 = \frac{2am^A\bar{\alpha}\sin^2(\tfrac{1}{2}\theta_p)}{\beta (a+\delta) \delta} \biggl( \text{ln} \, \biggl[ \frac{\tan(\tfrac{1}{2}\theta_p)}{\tan(\tfrac{1}{2}\theta_q)} \biggr] \biggr)^{-1} .
\label{eq:lambda}
\end{equation}
Finally, the speed of the colloid -- moving along the negative $z$-direction -- as well as the entire flow field outside the boundary layer may be obtained from Lighthill and Blake's analyses of squirming spheres~\cite{lighthill52,blake71}:
\begin{equation}
U = \lambda \sin^2(\tfrac{1}{2}\theta_p) ,
\label{eq:advective_speed}
\end{equation}
where we have neglected contributions from the radial slip that are smaller by a factor of ${\cal O}(\delta/a)$, and the fluid flow field outside the boundary layer is shown for a Janus particle ($\theta_p = \tfrac{\pi}{2}$) in Fig.~\ref{fig:flow}. 

Tacit in the foregoing analysis were the assumptions that the particle was acting as a source for the solute, $\alpha>0$, and that the interaction potential was repulsive, $V>0$. Reversing these signs independently provides four broad classes of self-diffusiophoretic motion, all of which lead to steady state motion at $\text{Pe}=0$~\cite{golestanian07}. However, at $\text{Pe}=\infty$ only two cases continue to support steady state propulsion; repulsive producers ($V > 0, \, \alpha > 0$) and attractive consumers ($V < 0, \, \alpha < 0$). The reason for this again lies in the advective transport through the boundary layer in which radial slip transports material away from, or towards, the surface according to whether the activity is positive (producing) or negative (consuming). By fluid continuity any such radial flux will be converted into a tangential one with $u_{\theta}^{\text{slip}}$ negative if there is an outflux and positive if there is an influx. Thus the particle motion will always be away from solute that it is producing, and towards solute that it is consuming, regardless of the interaction. But this motion is in the same direction as predicted by the boundary layer analysis of Eq. \eqref{eq:slip_general} only if the interaction is repulsive in the former case and attractive in the latter, so that only these cases provide consistent steady states at high Pe.  

The conditions for motility in the $\text{Pe} = \infty$ limit are therefore more restrictive than those at $\text{Pe} = 0$.   It is notable that the biological phenomena mentioned earlier, namely actin-driven propulsion of {\it Listeria} and chromosomal translocation in {\it Caulobacter} and {\it Vibrio}, appear to coincide with the two cases that lead to steady-state motility in the high Pe limit.  In {\it Caulobacter}, for example, the chromosome attaches to ParA via {\it parS} and ParB~\cite{shebelut}, suggesting that the interaction between the chromosome and ParA is attractive, so that the chromosome acts as an attractive consumer of ParA filaments~\cite{banigan11}.   In the case of {\it Listeria}, the bacterium surface and actin are negatively charged, so there should be a screened Coulomb repulsion between the two.  While the actin comet tail is attached to the bacterium, it is plausible that binding is sufficiently rare and transient~\cite{marchand,theriot} that there is a well-defined net interaction that is repulsive, making {\it Listeria} a repulsive producer of actin filaments.   We note that other mechanisms have been advanced for the case of {\it Listeria} motility~\cite{carlsson} in addition to self-diffusiophoresis~\cite{lee08}.  Since even at high Pe, asymmetric solute distributions give rise to motility with appropriate signs of interactions, the mechanism of self-diffusiophoresis must be reckoned with.

The speed \eqref{eq:advective_speed} found at high Pe should be compared with the corresponding expression  
\begin{equation}
U = \frac{m^D \bar{\alpha}}{4D} \sin^2(\theta_p) ,
\label{eq:diffusive_speed}
\end{equation}
for an identical active patch and interaction at low Pe~\cite{golestanian07}. Two observations are particularly noteworthy. 

First, the two limits exhibit distinct dependences on the size of the active patch $\theta_p$. In the diffusive regime the speed is invariant under $\theta_p \rightarrow \pi - \theta_p$ reflecting the fact that it is really the asymmetry that controls motility and, in particular, that the particle does not move if the coverage is uniform. At high Pe the situation is different, although some care is needed to interpret Eq. \eqref{eq:lambda} for the tangential flux constant $\lambda$. It is appropriate to rewrite the logarithmic term as $\text{ln}\, (a/\delta)$, which exhibits the relevant scaling without the artificial singularity associated with the neglect of the radial slip near $\theta = \pi$. With this mollifier the speed in the advective regime has the dependence $U \sim \sin^3 ( \tfrac{1}{2} \theta_p )$ and thus does not vanish in the limit $\theta_p \rightarrow \pi$ of complete coverage. One interpretation for this result is that a uniformly active particle, which has no natural directionality, is unstable in the limit ${\textrm Pe}=\infty$ with respect to a spontaneous symmetry breaking that selects a direction for motion. Such spontaneous symmetry breaking has been observed in the case of actin-driven motility if the colloid radius is small enough~\cite{cameron99,vandergucht05}. This qualitative agreement with the present analysis encourages a more thorough analysis of the onset of motility to explain the observed size dependence. 

Second, the two expressions exhibit different scalings with the activity rate $\bar{\alpha}$; in the diffusive regime, $U \sim \bar{\alpha}$, while in the advective regime, $U \sim \bar{\alpha}^{1/2}$. This discrepancy reflects precisely the different fluxes in Eq. \eqref{eq:solute_boundary_condition}. The diffusive flux is proportional to the tangential slip, whereas the advective flux is proportional to its square, and since the speed is proportional to the tangential flow the scalings follow. 
Recent experiments show that the propulsion speed of {\it Listeria} depends on the distribution of ActA, which catalyzes actin polymerization~\cite{rafelski}; the speed increases with the amount of ActA but it is difficult to adduce a functional form for the dependence of $U$ on $\bar \alpha$ from the data.

In summary, we have shown that objects that are repulsive producers or attractive consumers of non-diffusing solutes should propel themselves through a fluid.  Our abstract, minimal analysis suggests that in such situations, objects can spontaneously break symmetry to propel themselves in a given direction and achieve sustained steady-state motility.  Note that we have not taken into account the effect of the non-diffusing solute on the flow; for example, the actin comet tail of {\it Listeria} should behave as a porous medium to suppress flow within it.  Since this effect could be modeled with an inert trailing particle, which does not prevent motility at low Pe~\cite{kapral}, it seems unlikely to be a significant factor at high Pe, but this avenue should be pursued in future work.  Another important issue is the stall force needed to prevent motion.  At low Pe, the stall force is simply the drag force, but at high Pe it will be determined by the distortion of the concentration profile by the boundary layer flow. 

\acknowledgements{We gratefully acknowledge stimulating discussions with E. Banigan, T. Idema, R. Kamien and T. Lubensky.  This work was supported by UPENN-MRSEC DMR05-20020 (AJL and GPA) and DMR05-47230 (GPA).}


\begin{thebibliography}{99}
\bibitem{cameron99} L.A. Cameron, M.J. Footer, A. van Oudenaarden, and J.A. Theriot, Proc. Natl. Acad. Sci. USA {\bf 96}, 4908-4913 (1999). 
\bibitem{plastino05} J. Plastino and C. Sykes, Curr. Op. Cell Bio. {\bf 17}, 62-66 (2005). 
\bibitem{fogel06} M.A. Fogel and M.K. Waldor, Genes Dev. {\bf 20}, 3269-3282 (2006).
\bibitem{shebelut} C.W. Shebelut, J.M. Guberman, S. van Teefelen,  A.A. Yakhnina and Z. Gitai, Proc. Nat. Acad. Sci. {\bf 107}, 14194-14198 (2010).
\bibitem{paxton04} W.F. Paxton {\it et al.}, J. Am. Chem. Soc. {\bf 126}, 13424-13431 (2004).
\bibitem{golestanian05} R. Golestanian, T.B. Liverpool, and A. Ajdari, Phys. Rev. Lett. {\bf 94}, 220801 (2005). 
\bibitem{golestanian07} R. Golestanian, T.B. Liverpool, and A. Ajdari, New J. Phys. {\bf 9}, 126 (2007).  
\bibitem{howse07} J.R. Howse {\it et al.}, Phys. Rev. Lett. {\bf 99}, 048102 (2007). 
\bibitem{golestanian09} R. Golestanian, Phys. Rev. Lett. {\bf 102}, 188305 (2009).  
\bibitem{popescu09} M.N. Popescu, S. Dietrich, and G. Oshanin, J. Chem. Phys. {\bf 130}, 194702 (2009).
\bibitem{popescu10} M.N. Popescu, S. Dietrich, M. Tasinkevych, and J. Ralston, Eur. Phys. J. E {\bf 31}, 351-367 (2010).
\bibitem{sabass10} B. Sabass and U. Seifert, Phys. Rev. Lett. {\bf 105}, 218103 (2010). 
\bibitem{anderson89} J.L. Anderson, Ann. Rev. Fluid Mech. {\bf 21}, 61-99 (1989). 
\bibitem{lighthill52} M.J. Lighthill, Commun. Pure Appl. Math. {\bf 5}, 109-118 (1952).  
\bibitem{blake71} J.R. Blake, J. Fluid Mech. {\bf 46}, 199-208 (1971). 
\bibitem{banigan11} E.J. Banigan, M.A. Gelbart, Z. Gitai, N.S. Wingreen, and A.J. Liu, {\it submitted} (2011).
\bibitem{marchand} J.-B. Marchand, D.A. Kaiser, T.D. Pollard and H.N. Higgs, Nat. Cell Biol. {\bf 3}, 76-82 (2001).
\bibitem{theriot} M.J. Footer, J.K. Lyo and J.A. Theriot, J. Biol. Chem. {\bf 283} 23852-23862 (2008).
\bibitem{carlsson} A.E. Carlsson, in {\it Annual Review of Biophysics} (Eds. D.C. Rees, K.A. Dill, and J.R. Williamson) {\bf 39}, 91-110 (2010).
\bibitem{lee08} K-C. Lee and A.J. Liu, Biophys. J. {\bf 95}, 4529-4539 (2008). 
\bibitem{vandergucht05} J. van der Gucht, E. Paluch, J. Plastino, and C. Sykes, Proc. Natl. Acad. Sci. USA {\bf 101}, 7847-7852 (2005).
\bibitem{rafelski} S.M. Rafelski, J.B. Alberts and G.M. Odell, PLOS Comp. Biol. {\bf 5}, e1000434 (2009).
\bibitem{kapral} L.F. Valadares {\it et al.}, Small {\bf 6}, 565-572 (2010).
\end{thebibliography}
\end{document}